\documentclass[aps,pre, reprint]{revtex4-1}
\usepackage{graphicx}
\usepackage{hyperref}
\usepackage{xcolor}
\usepackage{amsmath}
\usepackage{amsfonts}
\usepackage{amssymb}
\usepackage{varioref}
\usepackage{float}
\usepackage{dcolumn}   % needed for some tables
\usepackage{subfigure}
\usepackage{array}
\usepackage{alphabeta}
\usepackage{mathtools}
\usepackage{commath}
\graphicspath{{plots/}}

\begin{document}

%\preprint{AIP/123-QED}

\title[]{Conformation and dynamics of a  self-avoiding  active flexible polymer}% Force line breaks with \\
%\thanks{Footnote to title of article.}

\author{Shalabh K. Anand}
\email{skanand@iiserb.ac.in}
% \affiliation[]{Department of Physics,\\ Indian Institute of Science Education and Research, \\Bhopal 462 066, Madhya Pradesh, India}%Lines break automatically or can be forced with \\
\author{Sunil P. Singh}%
 \email{spsingh@iiserb.ac.in}
\affiliation{ Department of Physics,\\ Indian Institute of Science Education and Research, Bhopal 462 066, Madhya Pradesh, India}%

%\author{C. Author}
% \homepage{http://www.Second.institution.edu/~Charlie.Author.}
%\affiliation{Second institution and/or address%\\This line break forced% with}%

%\date{\today}% It is always \today, today,  but any date may be explicitly specified

\begin{abstract}
We investigate conformations and dynamics of a polymer considering its monomers to be active Brownian particles. This active polymer shows very intriguing physical behavior which is absent in an active Rouse chain. The chain initially shrinks with active force, which starts swelling on further increase in force. The shrinkage followed by swelling is attributed purely to excluded-volume interactions among the monomers. 
In the swelling regime, chain shows a cross-over from the self-avoiding behavior to Rouse-behavior with scaling exponent $\nu_a\approx 1/2$ for end-to-end distance. The non-monotonicity in the structure is analysed through various physical quantities specifically,  radial distribution function of  monomers, scattering time, as well as various energy calculations. The  chain relaxes faster than the Rouse chain  in the intermediate force regime, with a cross-over  in variation of relaxation time  at large active force as given by a power-law $\tau_r\sim Pe^{-4/3}$ ($Pe$ is P{\'e}clet number).  
\end{abstract}
\pacs{Valid PACS appear here}

\maketitle

%\begin{quotation}
%The ``lead paragraph'' is encapsulated with the \LaTeX\ \verb+quotation+ environment and is formatted as a single paragraph before the first section heading. (The \verb+quotation+ environment reverts to its usual meaning after the first sectioning command.) Note that numbered references are allowed in the lead paragraph. The lead paragraph will only be found in an article being prepared for the journal \textit{Chaos}.
%\end{quotation}

%\section{Introduction}

\textit{Introduction\textemdash} A collection of freely moving active Brownian particles has drawn immense research activities in past few years in view of interdisciplinary  applications~\cite{vicsek1995novel,marchetti2013hydrodynamics,elgeti2015physics,bechinger2016active,toner1995long,dombrowski2004self}. These individual agents ballistically propel themselves by conversion of chemical energy into mechanical energy, thus their motion can be controlled in experiments in a desired manner consequently they display rich collective    dynamics~\cite{palacci2013living,singh2017non,arlt2018painting,lavergne2019group,lozano2016phototaxis,liebchen2018synthetic,yu2019phototaxis}. A collection of such active particles connected via linear chain exhibits numerous interesting  features ~\cite{biswas2017linking,nishiguchi2018flagellar,di2016active,yan2016reconfiguring,ashton2013self,nykypanchuk2008dna,chen2011directed,sasaki2014colloidal,ramirez2013polloidal,lowen2018active}, which is often absent in passive systems. For example, an active chain exhibits shrinkage and  swelling~\cite{anand2018structure,bianco2018globulelike,samanta2016chain,eisenstecken2016conformational,eisenstecken2017internal,osmanovic2017dynamics,martin2018active,shin2015facilitation}, spontaneous oscillations~\cite{chelakkot2014flagellar,laskar2013hydrodynamic,anand2019beating}, enhanced diffusion~\cite{anand2018structure,bianco2018globulelike,samanta2016chain,eisenstecken2016conformational,eisenstecken2017internal}, etc. The collective dynamics of such systems display various emergent structures, understanding them is a fundamental quest from biophysics point of view as it poses a great challenge~\cite{rodriguez2003conserved,vicsek2012collective,abkenar2013collective,decamp2015orientational,sumino2012large,schaller2013topological,ravichandran2019chronology,duman2018collective,prathyusha2018dynamically}.

%Fascinating and emergent behavior of active matter systems has been a subject of an immense interest in different disciplines of science~\cite{vicsek1995novel,marchetti2013hydrodynamics,elgeti2015physics,bechinger2016active} in last few decades.

%From the study point of view an active  polymers have been of interest from long time as  they intrinsically exhibit viscoelastic properties.  Because of the wide variety of bio-polymers in the nature, active polymers and rigid filaments in the presence of molecular motors are also of subject of research~\cite{winkler2017active,schaller2010polar,prost2015active,doostmohammadi2017onset,ndlec1997self,anand2019behavior,manna2019emergent,ganguly2012cytoplasmic,harada1987sliding,juelicher2007active,foglino2019non}. 
%They exhibit quite differently than their equilibrium counterpart. Understanding the properties of an active polymer is a fundamental quest form biological physics point of view. 

With the help of minimal models, behavior of an active flexible  chain or rigid filaments has been explored~\cite{winkler2017active,schaller2010polar,prost2015active,doostmohammadi2017onset,ndlec1997self,anand2019behavior,manna2019emergent,ganguly2012cytoplasmic,harada1987sliding,juelicher2007active,foglino2019non,ghosh2014dynamics,eisenstecken2016conformational,eisenstecken2017internal,samanta2016chain,peterson2019statistical,kaiser2014unusual,harder2014activity,kaiser2015does,anand2018structure,isele2015self,isele2016dynamics,laskar2017filament,jiang2014motion,das2019deviations,anand2019beating,bianco2018globulelike}. An accessible analytically tractable model for the polymer is Rouse model and inclusion of the activity in this model is studied in literature~\cite{eisenstecken2016conformational,eisenstecken2017internal,martin2018active,samanta2016chain,osmanovic2017dynamics,osmanovic2018properties,peterson2019statistical}. An active Rouse chain shows swelling with a power law scaling relation on active force with exponent $1/3$. Analytical calculations suggest that a flexible polymer always swells, whereas a semi-flexible chain shrinks at smaller force and in the asymptotic limit it swells akin to an active Rouse chain with same exponent~\cite{eisenstecken2016conformational,eisenstecken2017internal,martin2019active}. The swelling of chain, relaxation, and its centre-of-mass diffusion can be strongly influenced by the solvent properties and viscoelastic behavior of the medium~\cite{vandebroek2015dynamics,vandebroek2017generalized}. The competition between elastic and self-avoiding forces  causes shrinkage to passive chain in an active bath in two-spatial dimensions (2D)~\cite{harder2014activity}. On the other hand a self-avoiding active chain in 2D shrinks, which is followed by swelling at larger active strength~\cite{kaiser2015does}. How does excluded-volume interactions influence the structure of a chain in 3D, its relaxation, and scaling exponents  are important questions and have not been addressed very well in the previous studies. 
   
The present work elucidates the role of excluded volume together with the activity on the relaxation and structure of the chain. 
%The inclusion of multi-body excluded-volume interaction brings here a very interesting physics in structure of the chain which is absent in a Rouse chain~\cite{eisenstecken2016conformational,eisenstecken2017internal,samanta2016chain}.
 In our simulations, we found that the end-to-end distance ($R_{e}$) and radius of gyration ($R_{g}$) of the chain shrinks in the intermediate range of active force ($Pe$) in absence of hydrodynamics. In a recent study, it has been shown that the effect of hydrodynamics brings a similar behavior~\cite{martin2018active}. We analyse here the shrinkage of chain through relaxation time, mean collision of monomers, radial distribution function, softness of the potential, and  elastic and repulsive energies. The scaling exponent of the chain in stretching regime follows a power-law on active force as, $R_e\approx Pe^{1/3}$  and further with variation on the chain length as $R_e\approx N^{\nu_a}$, where $\nu_a \approx 1/2$ in the stretching regime. %We estimate the mean collision time here, which shows a non-monotonic behavior in the same range of P{\'e}clet number and this behavior of collision time changes by varying the excluded-volume parameter.  } 

%This article is organized as follows. Simulation model has been discussed in section II, whereas section III describes the outcomes and results. At the end we summarize the results and draw the conclusion of our study. }

%\section{Model}
\textit{Model\textemdash} A flexible chain is composed of linear sequence of $N$ Brownian particles, the consecutive monomers in the chain are connected by harmonic potential
%,connected via harmonic spring and excluded-volume interactions. 
%{\cred The harmonic and excluded-volume part of potential energy of the chain can be read as, $\Phi = \Phi_{h} + \Phi_{LJ}$, where $\Phi_{h}$ is spring potential, and  $\Phi_{LJ}$ corresponds to the excluded-volume interactions.} 
$\Phi_{h} = \frac{k_{s}}{2} \sum_{i=1}^{N-1}(|{\bm r}_{i+1} - {\bm r}_{i}|-l_{0})^2,$
 where ${\bm r}_i,l_0$, and $k_s$ denote position of $i^{th}$ monomer,  average equilibrium bond length and spring constant, respectively. The excluded-volume potential restricts overlapping of beads in a polymer, and it is implemented here as standard repulsive part of Lennard-Jones interactions for  shorter distance, {\it i.e.}, $R_{ij} < 2^{1/6}\sigma$,

\begin{equation}
u_{ij} = 
4 \epsilon \left[\left(\frac{\sigma}{R_{ij}}\right)^{12}- \left( \frac{\sigma}{R_{ij}}\right)^{6} \right] + \epsilon,
\end{equation}
 and for $R_{ij} \geq 2^{1/6}\sigma$, $u_{ij}=0$, where ${\bm R}_{ij}={\bm r}_j-{\bm r}_i$,  $\epsilon$ is interaction energy and $\sigma$ is the diameter of the monomer. The total LJ energy can be expressed as,  $\Phi_{LJ}=\sum_{i=1}^{N-1}\sum{'}_{j=i+1}^{N} u_{ij}$. The prime in second summation excludes the LJ interaction between consecutive bonded neighbors.

 The equation of motion of an active Brownian bead of the polymer chain in an overdamped limit is,
\begin{equation}
\gamma \frac{d {\bm r_i}}{d t} = - \nabla_{i} \Phi + \textbf{F}_{r}^{i} + F_{a}\hat{\textbf{u}}_{i},
\label{eq:langevin}
\end{equation}
where $\gamma$ is the friction coefficient, $\textbf{F}_{r}^{i}$ is the thermal noise with zero mean, and $F_{a}$ is the strength of self-propulsion force exerted on $i^{th}$ bead along $\hat{\textbf{u}}_{i}$ direction. The viscous drag and the thermal noise obey fluctuation-dissipation relation, $\left< \textbf{F}_{r}^{i}(t).\textbf{F}_{r}^{j}(t')
\right> = 6k_{B}T\gamma \delta_{ij}\delta(t-t')$. The long-range hydrodynamic interactions are neglected here. 

\begin{figure}%[h]
	\includegraphics[width=0.8\columnwidth]{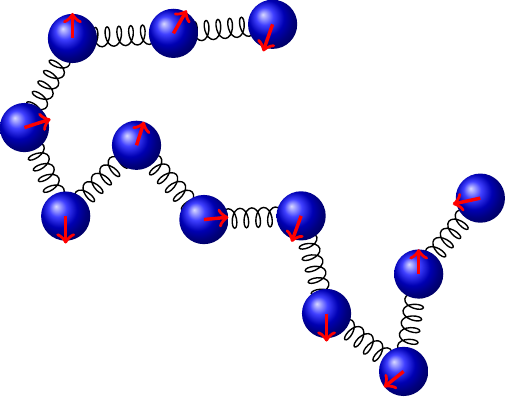}
	\caption{A pictorial snapshot of a modelled  polymer, arrow indicates the direction of active force on the corresponding monomer.}
	\label{Fig:model}
\end{figure}

Active Brownian beads are modelled as polar molecules, their orientations $\hat {\bm u}_i$ are described by the rotational counter-part of the Langevin equation,
\begin{equation}
\gamma_{r} \frac{d { \hat{\bm u}_i}}{d t} = {\bm \zeta_{i}} \times \hat{\textbf{u}}_{i}.
\label{eq:rotlangevin}
\end{equation}
Here ${\bm \zeta}_i$ is a random torque with zero-mean and variance $<{\bm \zeta}(t)\otimes{\bm \zeta}(t')> = 2(k_{B}T)^{2}\delta(t-t')/D_{r}$, and $\gamma_{r}$ is the rotational friction coefficient given as $\gamma_{r}=k_{B}T/D_{r}$. The rotational diffusion is expressed in terms of translational diffusion ($D_m$) as, $D_r={3D_m}/{l_0^2}$. The strength of active force is presented here as a ratio of active force with thermal force given as, $Pe = (F_{a} l_0)/(k_{B}T)$, with P{\'e}clet number $Pe$ as a dimensionless quantity. A schematics of polymer chain is displayed in Fig.~\ref{Fig:model}, where arrow shows the direction of active force on a monomer.

All the physical parameters presented in this article are scaled in units of the bond length $l_0$, diffusion coefficient of a monomer $D_{m}$, and thermal energy $k_{B}T$. Simulations are performed in cubic periodic boxes in three spatial dimensions, polymer length is varied in the range of $N=50$ to $300$. Other  parameters are chosen as, $k_{s}$ in range of $10^3$ to $10^4$ in units of $k_{B}T/l_{0}^{2}$, $\epsilon/k_{B}T=1$, and time is in units of $\tau = l_0^2/D_{m}$. For higher $Pe$, larger values of $k_s$ is chosen to avoid stretching of bonds. The monomer size $\sigma$ is varied in the range of $\sigma/l_0=0.2$ to $1.0$ and $Pe$ is varied in the range of $0$ to $1000$. We use Euler integration technique to solve Eq.~\ref{eq:langevin} and \ref{eq:rotlangevin} with time step $\Delta t$ in the range of $10^{-3}\tau$ to $10^{-5}\tau$ to ensure stable simulation results. In order to obtain better statistics, each data point is averaged over $20$ independent simulations.

%\section{Results}
\textit{Structural Properties\textemdash} There is vast literature on the equilibrium behavior of a polymer chain~\cite{winkler1994models,harnau1997influence,kratky1949diffuse,saito1967statistical,hsu2010polymer,hsu2011breakdown,doi1988theory,rubinstein2003polymer,muthukumar2016polymer} followed by extension to an active  chain~\cite{anand2018structure,isele2015self,man2019morphological}. We present effect of active noise on the structure of a self-avoiding chain in form of radius-of-gyration, end-to-end distance and its distribution, pair-correlation function, and scaling exponents. 
%\subsection{Structural Properties}

\begin{figure*}%[h]
	\includegraphics[width=0.65\columnwidth]{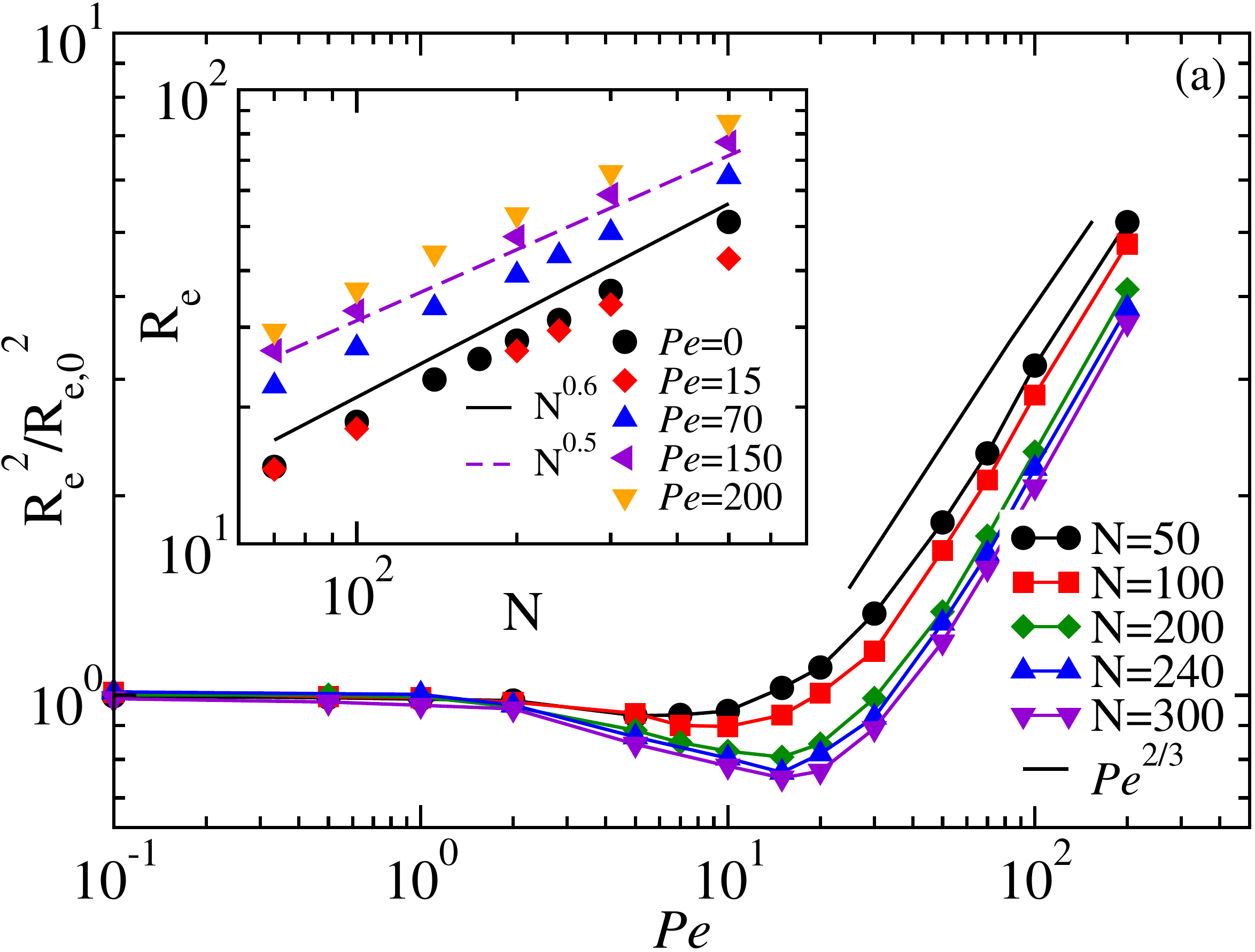}
	\includegraphics[width=0.65\columnwidth]{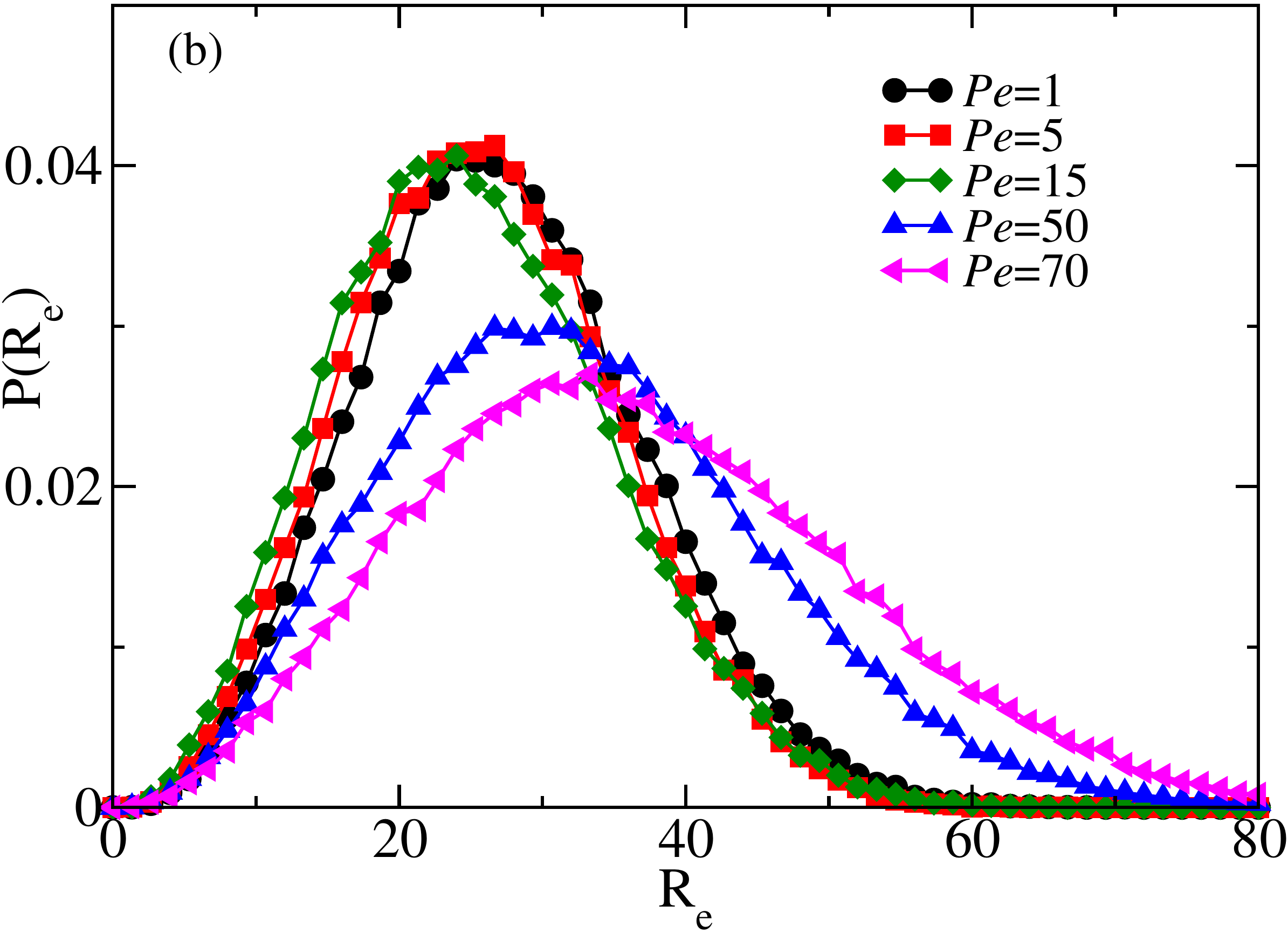}
	\includegraphics[width=0.65\columnwidth]{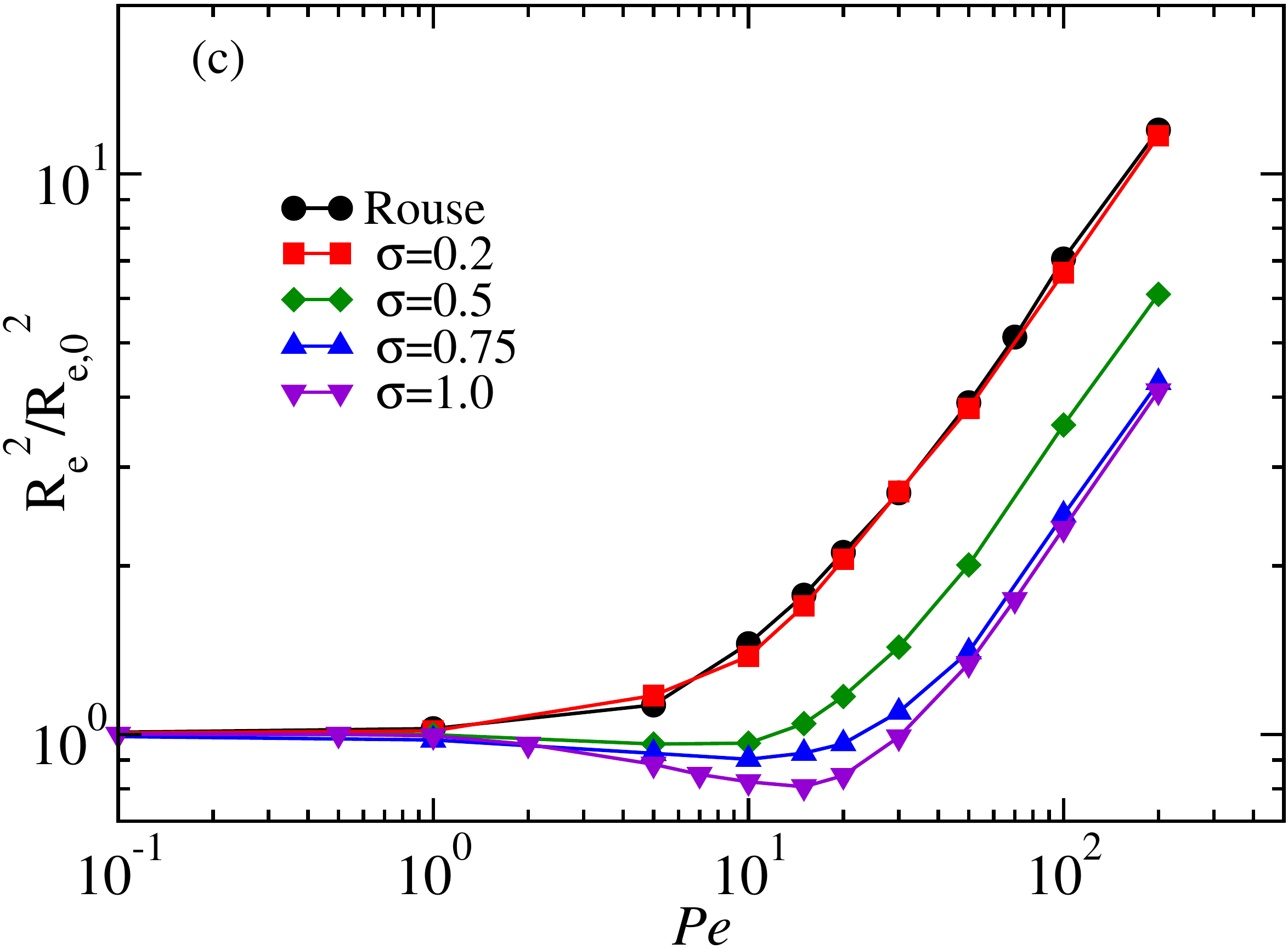}
	\caption{a) Relative variation of mean square end-to-end distance ($R_e^2/R_{e,0}^2$) of the chain as a function of $Pe$ for various lengths. Solid line shows a power law variation $Pe^{2/3}$. The inset shows $R_e$ with $N$ at $Pe=0,15,70,150$ and $200$, at $\sigma=1$. The solid and dashed lines are showing power law variation at exponents $\nu_a=3/5$ and $1/2$, respectively. b) The distribution of end-to-end distance at $N=200$ and $\sigma=1$. c) Relative variation of  end-to-end  distance ($R_e^2/R_{e,0}^2$) of an active chain  with $N=200$ as a function of $Pe$ for various monomer diameters $\sigma=0.2,0.5,.75$ and $1.0$, and Rouse chain (bullet).}
	\label{Fig:end_dst}
\end{figure*}
The  quantification of structural change is analysed in terms of end-to-end distance $R_{e}$ and radius of gyration $R_{g}$ as,
\begin{equation} R_e^2= \left<({\bm r_1}-{\bm r_N})^2\right>;~~R_g^2 = \frac{1}{N}\left<\sum_{i=1}^{N} ({\bm r}_i-{\bm R}_{cm})^2\right> ,
\end{equation}
where ${\bm R}_{cm}$ is the centre-of-mass of the chain and angular bracket indicates ensemble average. The computed $R_e$ is displayed   in Fig.~\ref{Fig:end_dst}-a, which reflects a significant shrinkage of polymer  with $Pe$ in the range of $Pe<50$ for $N>50$. 
The initial shrinkage of chain is followed by stretching in the range of $Pe>50$ as Fig.~\ref{Fig:end_dst}-a illustrates. The swelling behavior of $R_e$ appears quite alike to Rouse chain. The normalized end-to-end distance for various chain lengths follows the same trend with relatively higher compression for large chain lengths. 
In the stretching regime, $R_e$ follows a power-law variation on P{\'e}clet number given by $R_e^2\sim Pe^{2/3}$ with an exponent $2/3$ identical to Rouse chain~\cite{harder2014activity,ghosh2014dynamics,eisenstecken2016conformational}.

Now, we turn our attention to scaling exponents $\nu_a$ of the chain in various regimes. The inset of Fig.~\ref{Fig:end_dst}-a compares various plots of $R_e$ as a function of chain length at $Pe=0,15,70,150$, and $200$. These curves indicate  variation of the scaling exponents ${\nu_a}$ with $Pe$. It clearly suggest that for $1<Pe<50$, the exponent is slightly  smaller than $3/5$, for comparison a solid line is drawn in inset of Fig.~\ref{Fig:end_dst}-a at $\nu_a=3/5$. A dashed line  illustrates the variation of $Re\sim N^{\nu_a}$ with $\nu_a=1/2\pm0.05$. For $Pe>100$, the exponent $\nu_a$ of chain approaches Rouse regime $\nu_a=1/2$.  
To summarize the results in compression regime (triangle and diamond), we found that $\nu_a$ is smaller than $3/5$ and slightly larger than  $1/2$. 

The shrinkage of active chain is visible in terms of probability distribution of $R_e$. Figure~\ref{Fig:end_dst}-b reflects shift in the location of peak with propulsion strength $Pe$ at a fixed chain length $N=200$. The peak  shifts weakly  towards left for the smaller values of $Pe$ with shape almost identical to passive polymer. The initial shift of peak towards small $R_e$ changes its coarse of variation with shifting towards right for large $Pe$. The change in distribution is consistent with  the non-monotonicity in the structure. 
The end-to-end distance and probability distribution confirms the compression in the intermediate regime, i.e., $1< Pe < 50$. 

%\begin{figure}%[h]
%    \includegraphics[width=0.8\columnwidth]{pair_corr}
%	\includegraphics[width=0.8\columnwidth]{ener_per_part}
%	\caption{a) Radial distribution of the chain for various $Pe$ at $N=200$  and $\sigma=1$. b) Average excluded-volume energy per monomer as a function of $Pe$ for various chain lengths  at $\sigma=1$. Inset displays average elastic energy per monomer for various chain lengths as a function of $Pe$.}
%	\label{Fig:pair_corr}
%\end{figure}

\begin{figure}%[h]
	\includegraphics[width=0.8\columnwidth]{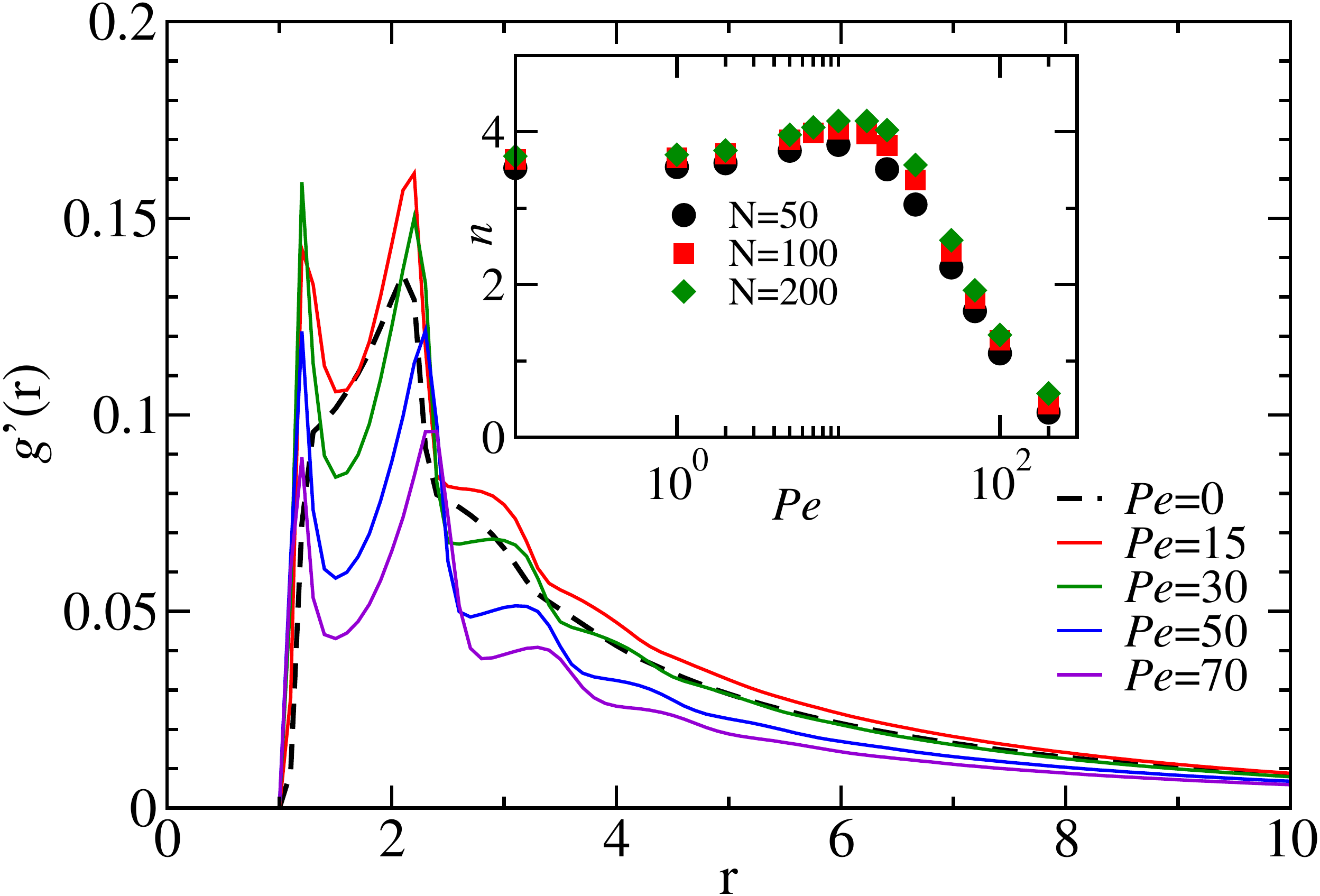}
	\caption{ Radial distribution of the chain for various $Pe$ at $N=200$, $\sigma=1$, and number density $\rho_0=7 \times 10^{-6}$. Here $g'(r)=\rho_0 g(r)$. Inset displays  average coordination number $n$  in first two shells ($R_{cut}=2.25$)  as function of $Pe$ for $N=50,100$ and $200$.}
	\label{Fig:pair_corr}
\end{figure}

In order to bridge the gap between monotonic swelling of an active Rouse chain and non-monotonic behavior of an excluded-volume chain, we quantify $R_e$ of chain by varying monomer's diameter $\sigma$. Figure~\ref{Fig:end_dst}-c illustrates normalized end-to-end distance $R_e^2/R_{e,0}^2$   with respect to its passive counterpart  $R_{e,0}^2$ as a function of $Pe$. Our simulations reveal that in the diameter range $0.2$ to $1.0$, we are able to find a smooth transition from Rouse to excluded-volume chain. For $\sigma=0.2$, $R_{e}$ displays a monotonic swelling with $Pe$ and the relative variations of $R_e$ are identical to  Rouse behavior. Further increase in $\sigma$  displays shrinkage in the chain. In the intermediate regime of $Pe$, the relative compression of the chain grows with $\sigma$ (see Fig.~\ref{Fig:end_dst}-c). The plot reveals  change in behavior from the continuous swelling regime to a shrinkage followed by swelling with variation in $\sigma$. This effect is attributed to increase in multi-body interactions with monomer diameter. This is discussed latter in the manuscript that shows how does active noise influence scattering time. 

The compression of chain indicates rise in local crowding in the intermediate regime $1<Pe<50$. To unveil this behavior, we estimate radial distribution function, which is a measure of average local density around a monomer. It is defined here as $g(r)=n(r)/(4\pi r^2 dr \rho_0 )$, here $n(r)$ is average number of monomer w.r.t. a given monomer in a concentric shell of radius $r$ and thickness $dr$. We have considered monomer density to be $\rho_0=7 \times 10^{-6}$, which is very small in dilute concentration of polymer, thus we present a scaled radial distribution $g'(r)=\rho_0 g(r)$ in the plot for better visualisation. Figure~\ref{Fig:pair_corr} displays radial distribution function of a chain, it clearly reveals a pronounced variation in the height of peaks  relative to passive chain. Hence, it shows higher local density in the intermediate $Pe$ regime, causing shrinkage of the chain. 
The height of peaks in distribution reverts its behavior for higher $Pe$ strengths and eventually becomes smaller than the passive chain. In this limit, the number of neighbors are less relative to passive chain as evident from its extension.    

%The role of rotational diffusion coefficient ($D_{r}$) becomes important here, as the monomers come together and stay together for some time. It takes $1/D_{r}$ time to change its orientation and escape from local crowded environment.
%Thus, the radial distribution reveals increase in local density due to excluded volume interactions in the intermediate range of P{\'e}clet number.
%it clearly reveals a pronounced variation in the height of peaks with activity. The magnitude of structured peaks grows at small separations with P{\'e}clet number in the intermediate regime (at $Pe=15$ and $30$). 
%The first and second peaks correspond to the probability of finding monomers in the first and second nearest neighbouring cells. 

\begin{figure}%[h]
	 \includegraphics[width=0.8\columnwidth]{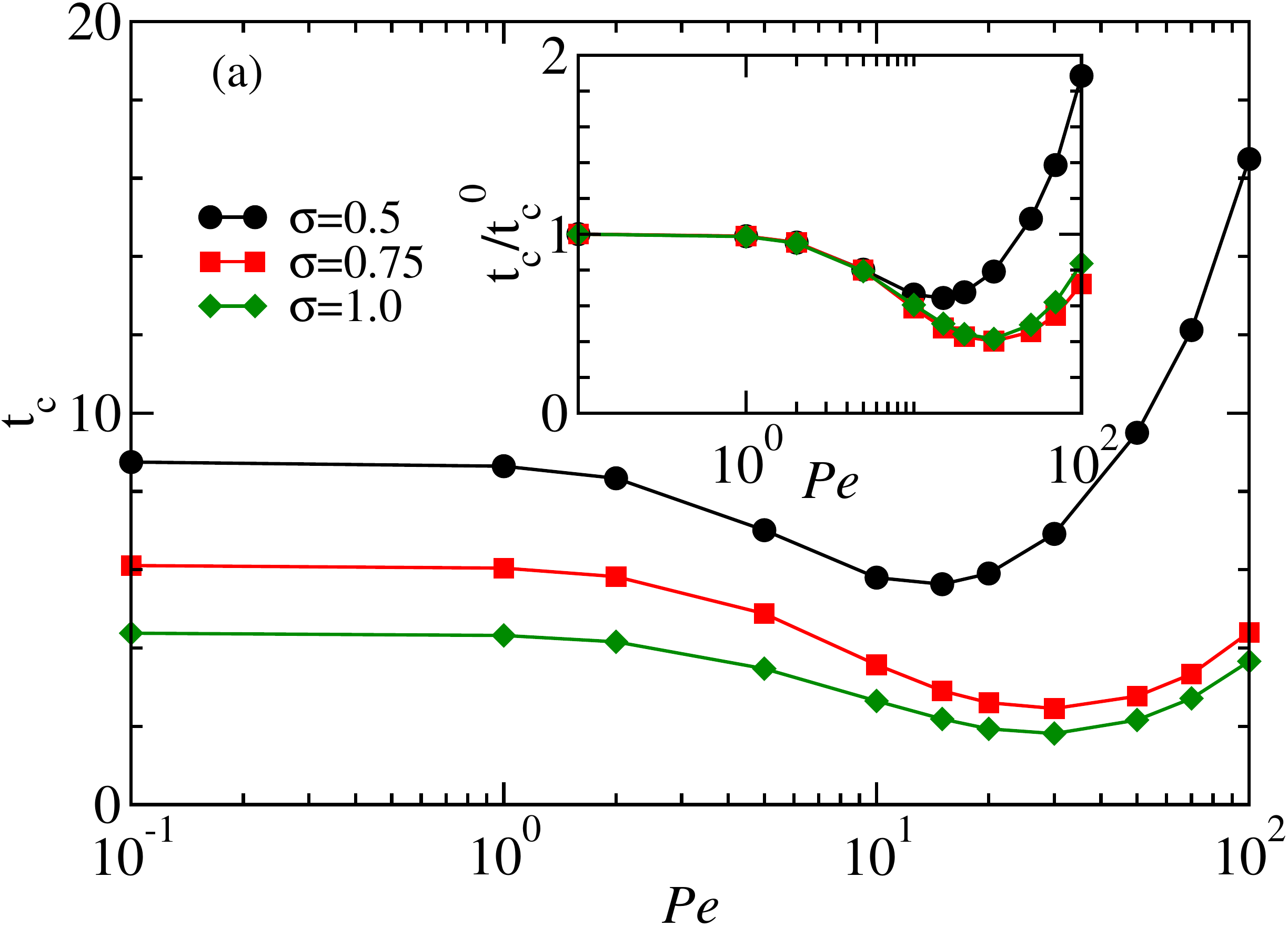}
	 \includegraphics[width=0.8\columnwidth]{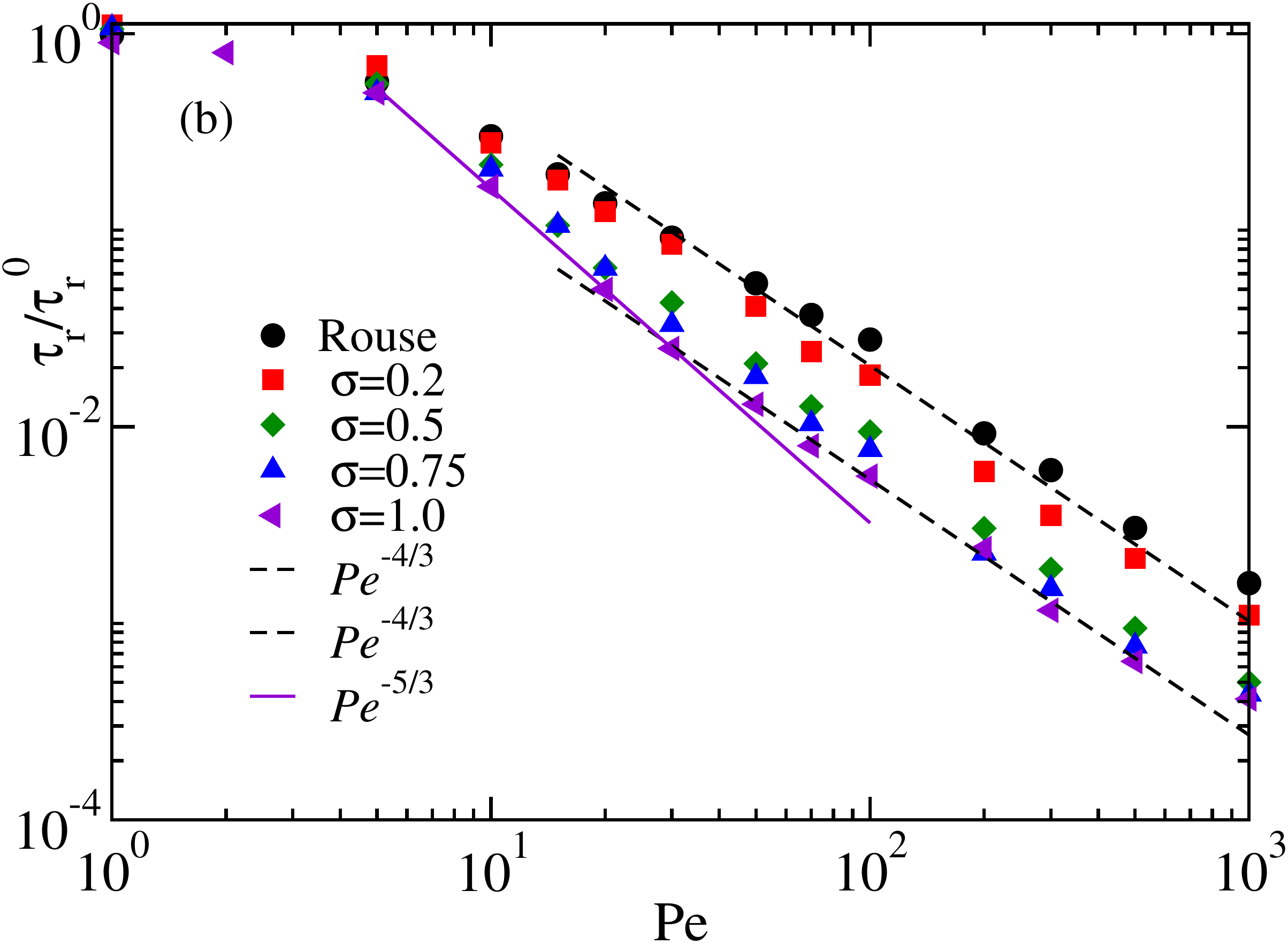}
	\caption{a) Average collision time $t_c$ of monomers  as a function of $Pe$ for $N=200$. Inset shows relative variation of $t_c/t_c^0$ with $Pe$, here $t_c^0$ is for $Pe=0$.  b) The relative variation of relaxation time $\tau_r/\tau_r^0$ for various monomer diameter $\sigma$ as a function of $Pe$ at $N=200$. The dashed  and solid lines show power law variation with exponent $4/3$ and $5/3$, respectively. }
	\label{Fig:collision}
\end{figure}

The variation in local coordination number in the intermediate range of P{\'e}clet number can be estimated from the radial distribution function as  $n=\int_{0}^{R_{cut}}4\pi r^2 g(r)dr$, where $R_{cut}$ is taken  up to the second peak at $Pe=0$ which is  $R_{cut}=2.25$. This gives the average coordination number  of chain in a cut-off distance ($R_{cut}$) as a function of $Pe$ (see inset of Fig.~\ref{Fig:pair_corr}). The local coordination number grows with $Pe$ in the cut-off distance, which indicates local accumulation of monomers and thus suggest shrinkage of  chain. In the  limit of $Pe>>1$, the local density  declines, thereby it signifies the stretching of chain.
%The effect of activity on the interaction of chain will be further illustrated in terms of variation in elastic and excluded volume energy contributions. The elastic energy per monomer is defined as $\phi_h=\Phi_h/N$ and excluded-volume energy $\phi_{LJ}=\Phi_{LJ}/N$. Figure~\ref{Fig:pair_corr}-b displays both energies as a function of $Pe$ for various chain lengths. Elastic energy is nearly unperturbed for small $Pe <10$, further it grows rapidly with $Pe$ in the range of $Pe>10$ as inset of Fig.\ref{Fig:pair_corr}-b illustrates. In the higher $Pe$ regime, monomers are randomly moving far apart with higher speed, which causes local stretching force on the chains thereby elastic energy grows. The contribution of the excluded-volume energies are displayed in Fig.~\ref{Fig:pair_corr}-b. As expected, $\phi_{LJ}$ exhibits sharp increase which is followed by a slump in the energy in the limit of $Pe>100$. The increase in $\phi_{LJ}$ indicates frequent collision of monomers, which is the effect of local crowding as a result higher energy. The LJ energy at sufficiently high $Pe>100$ decreases due to stretching of polymer resulting decrease in local density  thereby $\Phi_{LJ}$ decreases with $Pe$.

\begin{figure}%[h]
	\includegraphics[width=0.8\columnwidth]{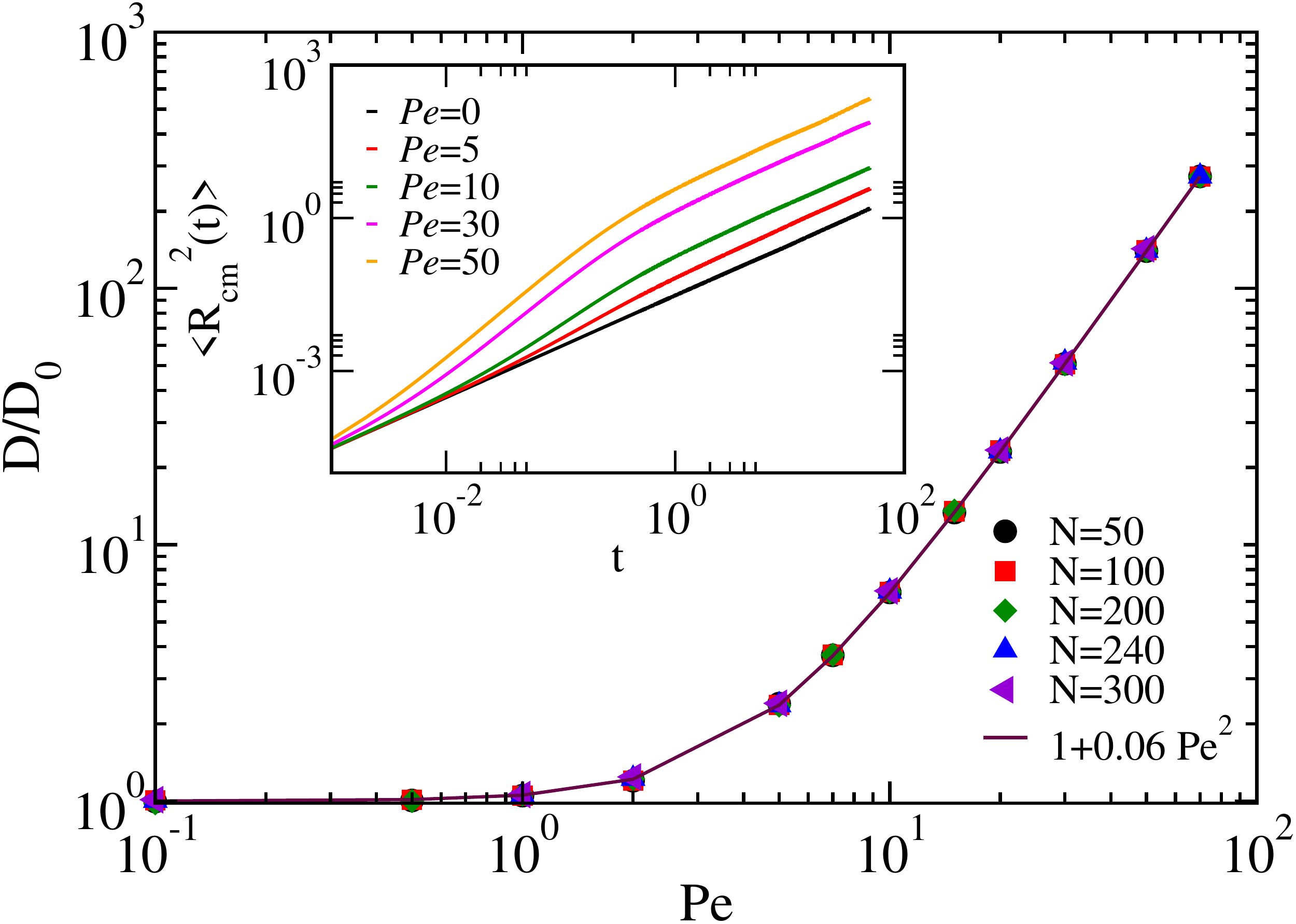}
	\caption{The effective diffusion coefficient  as a function of $Pe$ for various chain lengths. Inset shows mean squared displacement of the filament for chain length $N=200$ for various $Pe$ at $\sigma=1$. }
	\label{Fig:msd}
\end{figure}

%\subsection{Dynamics}
\textit{Dynamics\textemdash}
The understanding of non-monotonic behavior of active polymer's structure becomes more evident when we quantify average collision time $t_c$, and compare it with various monomer sizes. 
Overlap of two monomer's positions  in range of $R_c \le 2^{1/6}\sigma$ (LJ cut-off) is said to be a collision event. The average collision time provides a measure of hindrance or obstruction in motion of monomers in presence of self-avoidance.  Figure~\ref{Fig:collision}-a shows decrease in collision time $t_c$ followed by an increase with $Pe$ for  $\sigma=0.5,.75$ and $1.0$.  
 This is due to increase in local density and speed of  monomers with increasing activity in the range $0<Pe<50$. Further, they start dispersing far from each other on higher strength of $Pe$ as already pointed out  in terms of radial distribution. With higher strength ($Pe$), collision becomes frequent as expected from the kinetic theory $t_c \sim\frac{1}{v_r}$, $v_r$ read as average relative speed of monomers. The onset of increase of $t_c$ appears nearly at same $Pe$ as onset of $R_e$ and $R_g$ (see Fig.~SI-1-a in supplementary text). With increase in active fluctuations, the polymer gets stretched thus the frequency of collision goes down hence $t_c$ (collision time) goes up as Fig.~\ref{Fig:collision}-a reflects. The effect of monomer size on scattering time  indicates  variation in active polymer's conformation from self-avoiding to ideal behavior. A smaller monomer has larger collision time as it exhibits smaller scattering cross-section ($b=\pi \sigma^2$), which is reflected  in Fig.~\ref{Fig:collision}-a. It's noteworthy that the relative variation in $t_c/t_0$ for small $\sigma=0.5$ has strikingly significant variation, importantly in the intermediate regime of $Pe$ as inset of Fig.~\ref{Fig:collision}-a reflects. The depth in $t_c/t_0$ ($10<Pe<50$) becomes shallow with $\sigma$, which diminishes in the asymptotic limit of $\sigma \rightarrow 0$. 

To enlighten the difference in relaxation behavior of a Rouse and a self-avoiding active chain, we compute the end-to-end correlation of polymer. The end-to-end correlation follows an exponential decay (in longer time $\gamma_r t>>1$), $<\bm{R_{e}}(0).\bm{R_{e}}(t)> \simeq \exp(-t/\tau_{r})$, with $\tau_{r}$ as the longest relaxation time of polymer. The estimated relaxation time $\tau_r/\tau_r^0$ from the correlation is displayed in Fig.~\ref{Fig:collision}-b. It presents relative variation of $\tau_r$ with respect to that of passive chain $\tau_r^0$ for various monomer diameters, along with Rouse chain.  In the limit of smaller monomer size, we achieve relaxation behavior of Rouse chain with pronounced variation in $\tau_r$. The relaxation behavior indicates a power-law variation given as, $\tau_r\approx Pe^{-\beta_a}$, for Rouse chain  $\beta_a\approx 4/3$. The self-avoiding chain exhibits very intriguing feature  with a sharp variation $\tau_r$ in the limit $Pe<100$. The scaling  exponent is found to be $\beta_a\approx 5/3$, see solid line in  Fig.\ref{Fig:collision}-b. A cross over from the sharp relative variation ($\beta_a\approx 5/3$ for $Pe<100$) to the exponent $\beta_a=4/3$ is observed in the limit of $Pe>100$.  More importantly the variation in relaxation with $Pe$ becomes faster in compression regime.  The larger relative change in the relaxation time in the presence of hydrodynamics than the Rouse chain is also reported in Ref.~\cite{martin2019active}, where it was shown  that the competition between active force and variation in $\tau_{r}$ controls the structure  leading to compression of chain. The relatively faster variation of $\tau_{r}$ obtained in our simulations for self-avoiding chain resembles results of Ref.~\cite{martin2019active}. A smooth variation in $\tau_r$ from the self-avoiding chain to Rouse chain is obtained by  variation in monomer size as Fig.~\ref{Fig:collision}-b  illustrates in  various plots.

In this section, diffusion of active chain is presented through mean squared displacement (MSD) of the centre-of-mass (COM) of chain. The long time MSD of the COM is expressed as, $<R_{cm}(t)^{2}> = <(r_{cm}(t)-r_{cm}(0))^2>$. The MSD  shows ballistic motion $<R_{cm}^{2}(t)> \sim t^{2}$ in the short time and diffusive regime $<R_{cm}^{2}(t)>=6Dt$ in the long time limit. Inset of  Fig.~\ref{Fig:msd} displays MSD of the chain at various $Pe=0,5,10,30$ and $50$, it clearly indicates  enhanced diffusion with $Pe$. This can be understood in terms of  drag of monomers through active forces in random directions, which causes faster movement relative to the passive monomers  resulting in enhanced MSD of the chain with $Pe$. The diffusive regime of the MSD gives self-diffusion coefficient, as Fig.~\ref{Fig:msd} displays normalized effective diffusion coefficient $D/D_0$. The effective diffusivity increases quadratically as, $D \sim Pe^{2}$. Moreover as expected, $D$ is independent of the chain length when scaled by  diffusion coefficient of passive chain ($D_0$, at $Pe=0$),  thus we can express,
$D = D_{0}(1+aPe^{2})$, 
where $a\approx0.06$ is a constant. The effective diffusion can be used to define the effective temperature of the chain as $T_{eff}=1+aPe^{2}$. In particular cases, this expression is argued to be identical to a passive system with temperature equivalent to $T_{eff}$~\cite{palacci2010sedimentation,loi2011effective,loi2011non}.  However, mapping of effective temperature of active polymer to temperature would not be sufficient for all physical behaviors.

The segmental MSD of the chain reveals internal dynamics specifically sub-diffusive behavior in the intermediate time limit ($10^{-1}$ to $10^2$). The cross-over from sub-diffusive to diffusive survives relatively at longer time for larger chain lengths in a broad window of $Pe$ (see ESI Fig.~SI-2 a and b). This enlightens the internal dynamical picture of  chain in the discussed parameter space. 
%This indicates the restricted motion responsible for the sub-diffusion persist for longer time before its orientation gets looses memory. 

%\section{Discussion and Summary}
\textit{Discussion and Summary\textemdash}
In summary, we have unveiled the effect of excluded-volume interactions on the structural properties and internal dynamics of an active polymer in 3D.  A polymer shrinks in the presence of activity, which is followed by swelling. A pronounced non-monotonic behavior in $R_{e}$ is depicted in a broad range of activity strength  for larger chain lengths. This compression is more pronounced in 2D~\cite{kaiser2014unusual}. We have shown that in limit of  $Pe<50$, the compression is primarily a consequence of interaction of monomers from its neighbors, which brings an increase in local density. 
%{\cred The radial distribution function substantiate the effect of softness and increase in local density. The softness of repulsive potential exhibits a weak contribution in the structure at large active forces.  A systematic study on the softness of potential due to activity and the influence of soft  potentials on the structure of an active chain would be further interesting to investigate. }
This increase can be understood in terms of rotational diffusion which requires   $1/D_{r}$ time to change monomers orientation to escape from  the local environment.
The radial distribution function substantiates the effect of softness and increase in local density. The softness of repulsive potential exhibits a weak contribution in the structure at large active forces.  A systematic study on the softness of the potential due to activity  is taken into account here by varying $\epsilon$  over a range of $10$ to $10^{-3}$. Larger epsilon corresponds to stiffer potential, which exhibits very nominal change in the values with preserving the  qualitative behavior. On the otherhand  relatively softer potentials lead to a significant change in $R_e^2$ with activity (see Fig-SI 3).  We have also tested our results for a different potential which looks similar in nature to LJ potential but more steeper. This potential also exhibits a non-monotonic behavior in structure.
 
 Fast random motion of monomers results into stretching of the chain for large $Pe>50$, thereby increase in the elastic energy. Interestingly, the power law scaling exponent of self-avoiding chain ($R_e\approx N^{\nu_a}$)  $\nu_a$ becomes smaller in this regime and approaches to Rouse limit  ($\nu_a=1/2$), despite stretching of polymer due to activity. In addition, the longest relaxation decreases with power law as $\tau_r\approx Pe^{-\beta_a}$, for Rouse and self-avoiding chains in the large $Pe$ limit with exponent $\beta_a\approx4/3$. The relaxation behavior of self-avoiding chain's exhibits a crossover from the exponent $\beta_a\approx 5/3$ to $4/3$ with $Pe$.  
 In conclusion, the role of self-avoidance has been explored in a systematic way by varying monomer's diameter that bridges the gap between an excluded volume chain and a Rouse chain, consequently it connects variation of numerous physical properties such as $R_e$, $\tau_r$, and scattering time smoothly from one to another limit. 
 The effect of excluded volume is substantial in  flexible limit, which slowly  diminishes with  semi-flexibility of the chain \cite{eisenstecken2017conformational}.
 A theoretical approach for the radius of gyration and  relaxation time of the excluded volume chain would be essential for the complete understanding of system. In addition  a detailed study on the softness of potential and effect of rotational diffusion on the structure of an active chain would be further interesting to investigate.

\textit{Acknowledgements\textemdash} Authors acknowledge HPC facility at IISER Bhopal for the computation time. SPS thanks DST SERB Grant No. YSS/2015/000230 for the financial support. SKA thanks IISER Bhopal for the funding. Authors thank M. Muthukumar and Roland G. Winkler for indulgence in various useful and insightful discussions on the subject.

%\nocite{*}
%\bibliographystyle{apsrev}
%\bibliography{references}

 \pagebreak
 \renewcommand{\thefigure}{SI-\arabic{figure}}
 \setcounter{figure}{0}
 \section*{Supplementary text}

\begin{figure}%[h]
	\includegraphics[width=\columnwidth]{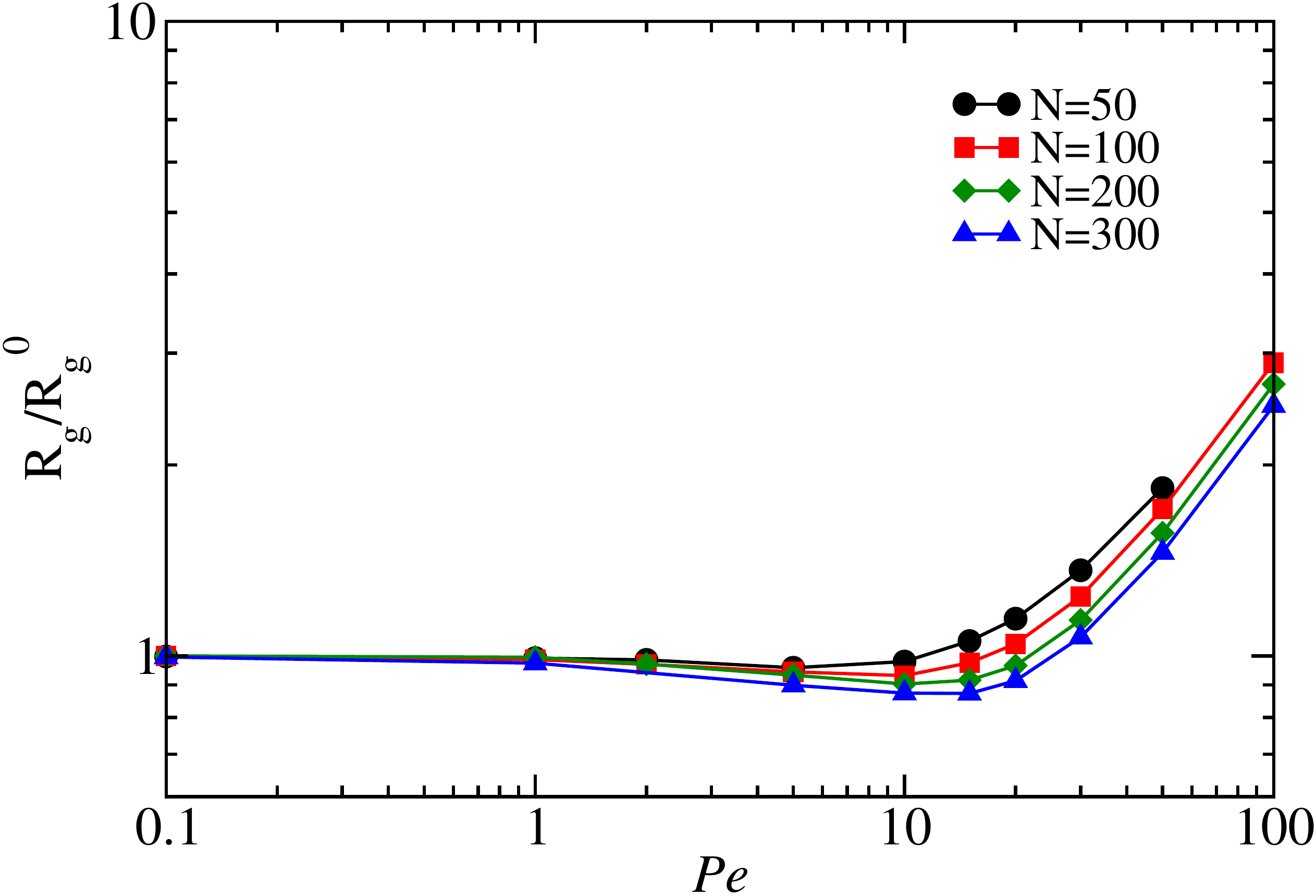}
	\includegraphics[width=\columnwidth]{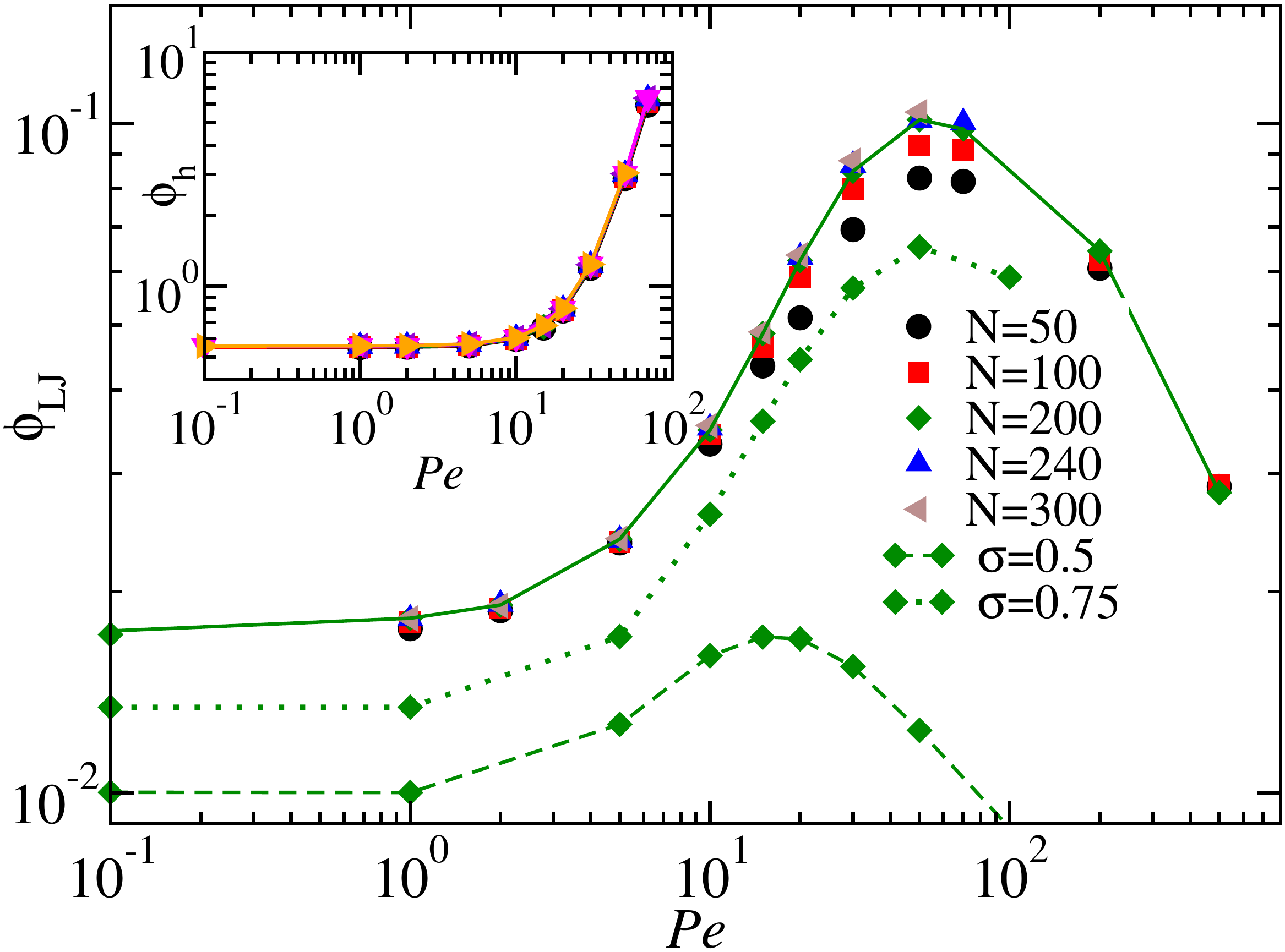}
	\caption{a) Radius of gyration of the chain as a function of $Pe$ for various chain lengths. b) Average excluded-volume energy per monomer as a function of $Pe$ for various chain lengths  at $\sigma=1$. Inset displays average elastic energy per monomer for various chain lengths as a function of $Pe$.}
	\label{Fig:SI rad_gyr}
\end{figure}

\begin{figure}%[h]
	\includegraphics[width=\columnwidth]{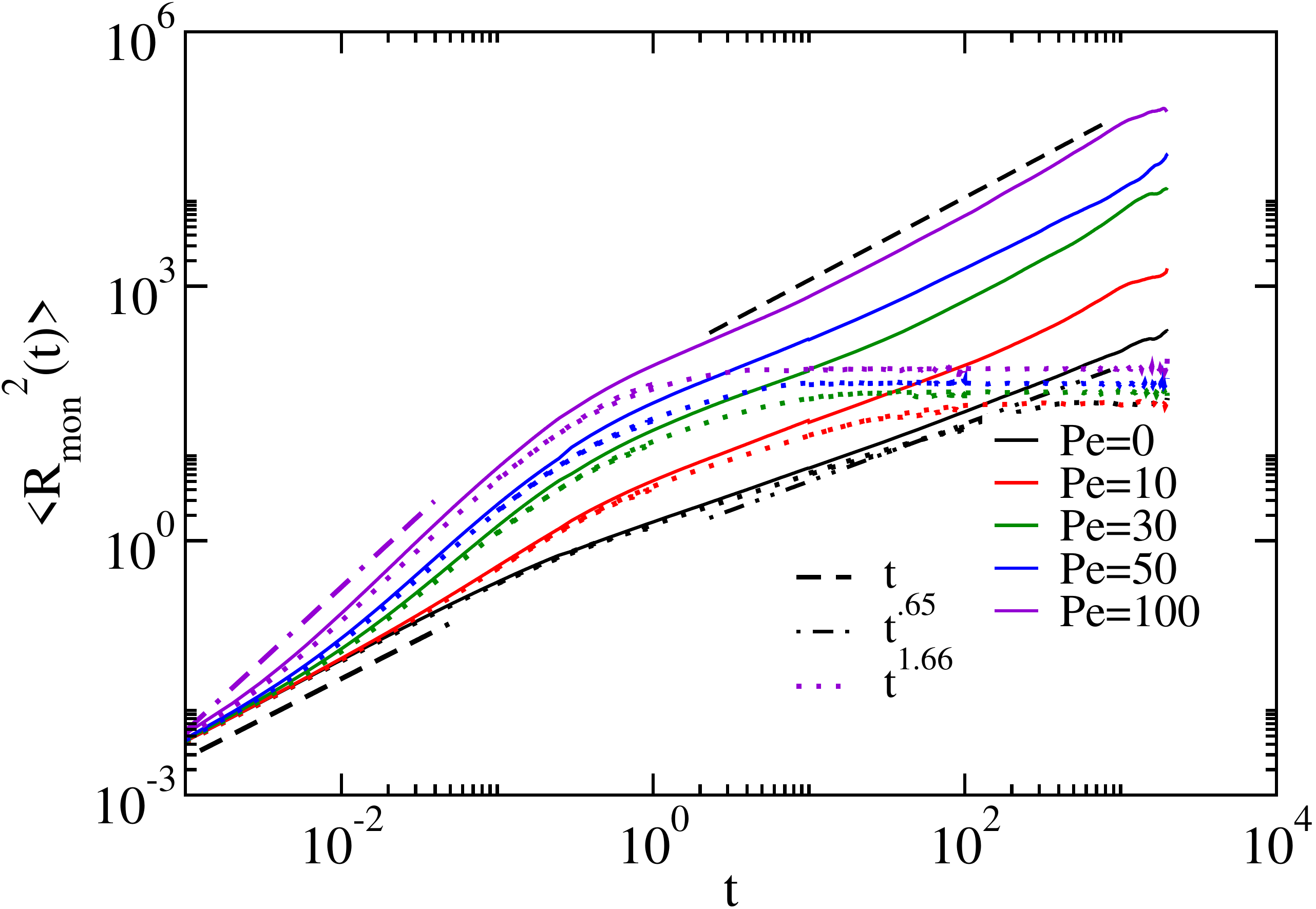}
	\includegraphics[width=\columnwidth]{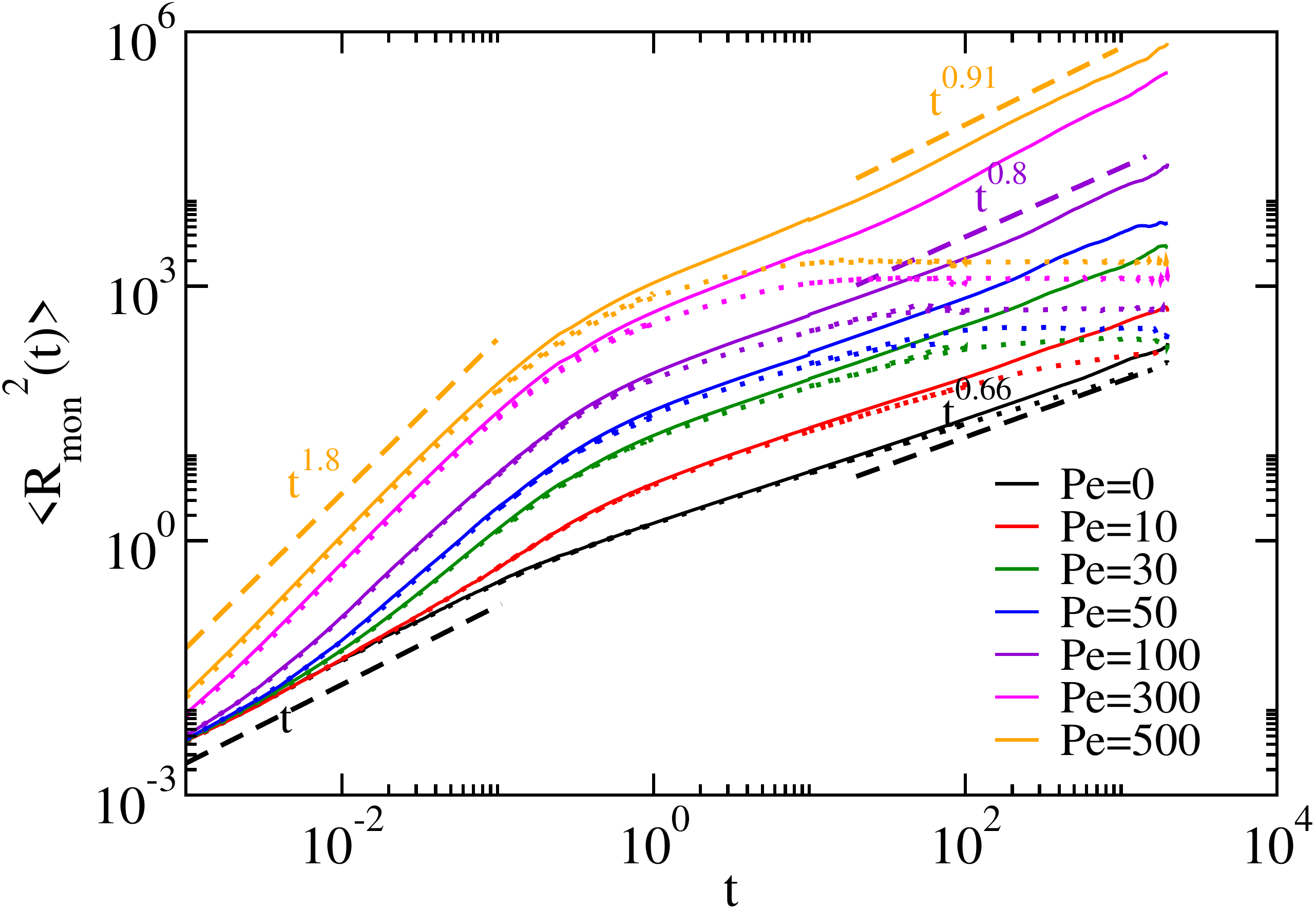}
	\caption{Mean squared displacement of monomers for chain-length a) $N=50$ and b) $N=200$ at $\sigma=1.0$. The dashed lines in both plots represent  MSD of monomer w.r.t. COM of the chain.}
	\label{Fig:SI mon msd}
\end{figure}
The Fig.~\ref{Fig:SI rad_gyr}(a) displays the radius of gyration of chain w.r.t. $Pe$. It shows a non-monotonic behavior for various chain lengths. 
% This also compares the radius of gyration for the a different potentials which is much steeper than the LJ potential. This shows that the radius of gyration is nearly unperturbed and it shows a weak variation with the effective temperature.}  

The effect of activity on the interaction of the chain is also illustrated in terms of variation in elastic and excluded volume energy contributions. The elastic energy per monomer is defined here as $\phi_h=\Phi_h/N$ and excluded-volume energy as $\phi_{LJ}=\Phi_{LJ}/N$. Figure~\ref{Fig:SI rad_gyr} (b) displays both energies as a function of $Pe$ for various chain lengths. The elastic energy is nearly unperturbed for  $Pe <10$,  it grows further rapidly with $Pe$ in the range of $Pe>10$ as inset of Fig.\ref{Fig:SI rad_gyr} (b) illustrates. In the higher $Pe$ regime, monomers are randomly moving  with relatively larger speed, which causes local stretching on the chain thereby elastic energy grows. The contribution of the excluded-volume energy is displayed in Fig.~\ref{Fig:SI rad_gyr} (b). As expected, $\phi_{LJ}$ exhibits sharp increase  followed by a slump in the energy in the limit of $Pe>100$. The increase in $\phi_{LJ}$ indicates frequent collision of monomers, which is the effect of local crowding as a result higher $\phi_{LJ}$. The LJ energy at sufficiently high $Pe>100$ decreases due to stretching of polymer resulting decrease in local density. 

The excluded volume energy  for various $\sigma=1.0,0.75$ and $0.5$ is also estimated. As expected, average excluded volume energy grows with monomer size and it also grows with $Pe$ in the intermediate regime and diminishes in the large limit of $Pe$. The height of peak decreases for smaller $\sigma$, and $\phi_{LJ}$ will approach zero for  very small $\sigma$.

The MSD of a monomer in chain is displayed ($N=50$ and $200$)  in Fig.~\ref{Fig:SI mon msd}(a) and (b). As evident from the figure, the MSD of a monomer of a passive chain undergoes diffusive, sub-diffusive followed by a long-time diffusive behavior. In the intermediate time $10^{-1}$ to $10^2$, the MSD exhibits sub-diffusive motion with exponent $t^{\alpha}, \alpha\approx2/3$.    
At moderate strength of $Pe$, a super-diffusive motion at shorter times, and sub-diffusive at relatively longer times is shown in the plot. The time window of the sub-diffusive regime narrows with $Pe$, and it almost disappears for shorter chain for $Pe=50$ and $100$. However, the sub-diffusion for relatively longer chain, i.e., $N=200$, persist in the active regime of $Pe<100$ and  it disappears beyond $Pe>100$. Thus for longer active chain, motion of monomers are influenced by other monomers for much longer time.

\begin{figure}%[h]
	\includegraphics[width=\columnwidth]{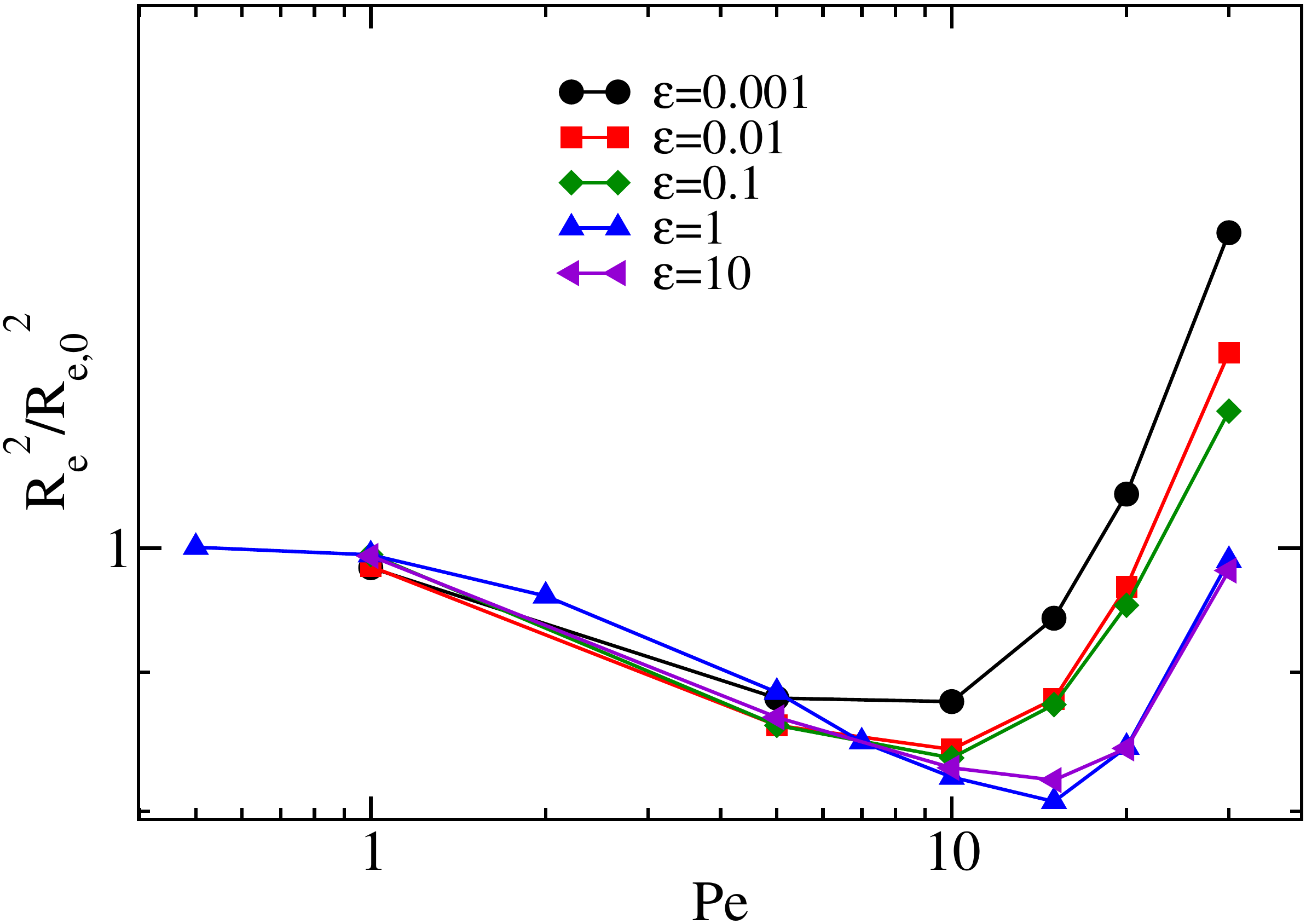}
	\caption{End-to-end distance of the chain as a function of $Pe$ for different values of LJ energies ($\epsilon$).}
	\label{Fig:SI_re}
\end{figure}

%The non-monotonic behavior of the results are  consistent for other potentails which is much sharper in nature. For example, we quantify $R_{e}^{2}$ of the chain for the potential which is of the given form,
%\begin{equation}
%u_{ij} = 
%4 \epsilon \left[\left(\frac{\sigma}{R_{ij}}\right)^{48}- \left( \frac{\sigma}{R_{ij}}\right)^{24} \right] + \epsilon.
%\end{equation}
%The parameters of the above potential is chosen in way that it exhibits same radius of gyration as LJ polymer for the passive case.  
%Figure~\ref{Fig:SI re} displays the non-monotonic behavior in structure similar to standard LJ potential. Here relative compression is smaller than than the LJ case, but the qualitative nature of the non-monotonic case is same. 

A systematic study of end-to-end distance of the active chain due to  softness of potential is taken  into account here by varying $\epsilon$ over a range of $10$ to $10^{-3}$. Here larger epsilon  corresponds to relatively steeper potential. Figure.\ref{Fig:SI_re} compares results of various $\epsilon$, it  exhibits very nominal change in the values of $R^2_e$ for $\epsilon=1$ and $10$. However for the $\epsilon<1$, the  potentials becomes softer thus the relative compression of the chain decreases. In the limit of $\epsilon \rightarrow 0$ the results will approach to the Rouse behaviour.

\end{document}